\input{aipcheck}

\documentclass[
    ,final            
  ]
  {aipproc}

\usepackage{amsmath}
\usepackage{graphicx}
\usepackage{epstopdf}
\usepackage{bm}
\usepackage{epsfig}
\usepackage{graphics}
\usepackage{xspace}
\usepackage{pstricks}
\usepackage{subfigure}

\layoutstyle{8x11single}

\def\Dslash{D\!\!\!\!\slash}

\newcommand{\nn}{\nonumber}

\newcommand{\bea}{\begin{eqnarray}}
\newcommand{\eea}{\end{eqnarray}}

\newcommand{\be}{\begin{equation}}
\newcommand{\ee}{\end{equation}}

\begin{document}

\title{Probing Long Range Scalar Dark Forces in Terrestrial Experiments}


\author{Sonny Mantry\footnote{Contributed talk at the Invisible Universe Conference, 2009, Paris, France.}}{
  address={University of Wisconsin at Madison,WI,53705,USA}
}

\begin{abstract}
A long range Weak Equivalence Principle  (WEP) violating force between Dark Matter  (DM) particles, mediated by an ultralight scalar, is tightly constrained by galactic dynamics and large scale structure formation. We examine the implications of such a \lq\lq dark force" for  several terrestrial experiments, including  E\"otv\"os tests of the WEP and  direct-detection DM searches. The presence of a dark force implies a non-vanishing effect in E\"otv\"os tests that could be probed by current and future experiments depending on the DM model. For scalar singlet DM scenarios, a dark force of astrophysically relevant magnitude is ruled out  in large regions of parameter space by the DM relic density and WEP constraints. WEP tests also imply constraints on the Higgs-exchange contributions to the spin-independent (SI) DM-nucleus direct detection cross-section. For WIMP scenarios, these considerations constrain Higgs-exchange contributions to the SI cross-section to be subleading compared to gauge-boson mediated contributions. The
combination of observations from galactic dynamics, large scale structure formation, E\"otv\"os experiments, DM-direct-detection experiments, and colliders can  further constrain the size of new long range forces in the dark sector.

\end{abstract}

\maketitle


\section{Introduction}
Despite the large body of evidence for Dark Matter (DM) from the galactic rotation curves~\cite{Bosma:1981zz, Rubin:1985ze}, acoustic 
oscillations in the cosmic microwave background~\cite{Hu:1994uz,Hu:1994jd, Jungman:1995bz,Zaldarriaga:1997ch}, large scale structure formation~\cite{Eisenstein:1997ik, Eisenstein:2005su}, and gravitational lensing~\cite{Clowe:2006eq, Zhang:2007nk} almost nothing is known about the properties of DM and a great deal of experimental effort is under way to unravel its nature.   Ground based direct detection experiments~\cite{Angle:2007uj,Ahmed:2008eu} put limits on the DM mass and the strength of its interaction with baryonic matter from observations of recoiling nuclei.  Experiments~\cite{Adriani:2008zr, Barwick:1997ig,Beatty:2004cy,Aguilar:2007yf} studying cosmic rays from the galactic halo 
have recently seen indications of an electron/positron excess, which could be interpreted as evidence for DM annihilation, and can constrain the DM mass and interactions. 

Another set of experiments are devoted to question of whether DM violates the Weak Equivalence Principle (WEP). Such an apparent violation of the WEP can occur in the presence of a new long range scalar force coupled to DM and is the focus of this work. In particular, we consider the possibility of an ultralight scalar with mass $m_\phi < 10^{-25}$ eV coupling directly to DM with intergalactic range. There are a few hints for the presence of a dark force from the  higher than predicted supercluster densities \cite{Einasto:2006si},  and voids~\cite{Farrar:2003uw, Nusser:2004qu}, and galaxy clusters~\cite{Baldi:2008ay} ( for a summary see \cite{Kesden:2006vz,Bovy:2008gh}).  Strong constraints on a dark  force are derived from observations of DM dynamics in the tidal stream  of the Sagittarius dwarf galaxy~\cite{Kesden:2006vz,Kesden:2006zb}, which indicate a force with strength less that 20\% of gravity.  However, new observational systematic errors have been recently discovered~\cite{Chou:2006ia} that could require a revision of this result, perhaps allowing for a stronger dark force. A more recent analysis~\cite{Bean:2008ac} considers the effect of a dark force on the evolution of density perturbations and the resulting impact on the CMB spectrum with similar constraints.

Many  models that contain the interaction of an ultralight scalar with DM~\cite{Damour:1990tw,Friedman:1991dj, Gradwohl:1992ue,  Anderson:1997un, Carroll:1998zi, Amendola:2001rc, Farrar:2003uw, Gubser:2004du, Gubser:2004uh, Bertolami:2004nh, Nusser:2004qu, Bean:2007ny} have been proposed to explain features in the DM distribution and explore the possibility of DM-quintessence interactions. More recently, work with non-universal scalar-tensor theories of gravity with the Abnormally Weighting Energy (AWE) Hypothesis~\cite{Alimi:2008ee, Fuzfa:2007sv} also invoke couplings of an ultralight scalar to the dark sector as a way of explaining the observed cosmic acceleration even in the absence of a dark energy fluid. Constraints on such scenarios from big bang nucleosynthesis have also been studied~\cite{Coc:2008yu}. 

If the DM has interactions with the Standard Model (SM), the presence of a scalar dark force will be communicated to the SM via virtual DM. This effect can be exploited to connect signatures of dark forces in astrophysics and cosmology with terrestrial experiments such as the laboratory E\"otv\"os tests of the WEP, DM-direct-detection  experiments, and in some cases even at colliders.
This connection has been studied in \cite{Bovy:2008gh}  and \cite{Carroll:2008ub} and more recently an extensive analysis was done in ~\cite{Carroll:2009dw} and is the focus of this note.
\begin{figure}
\includegraphics[height=2.1in, width= 2.5in]{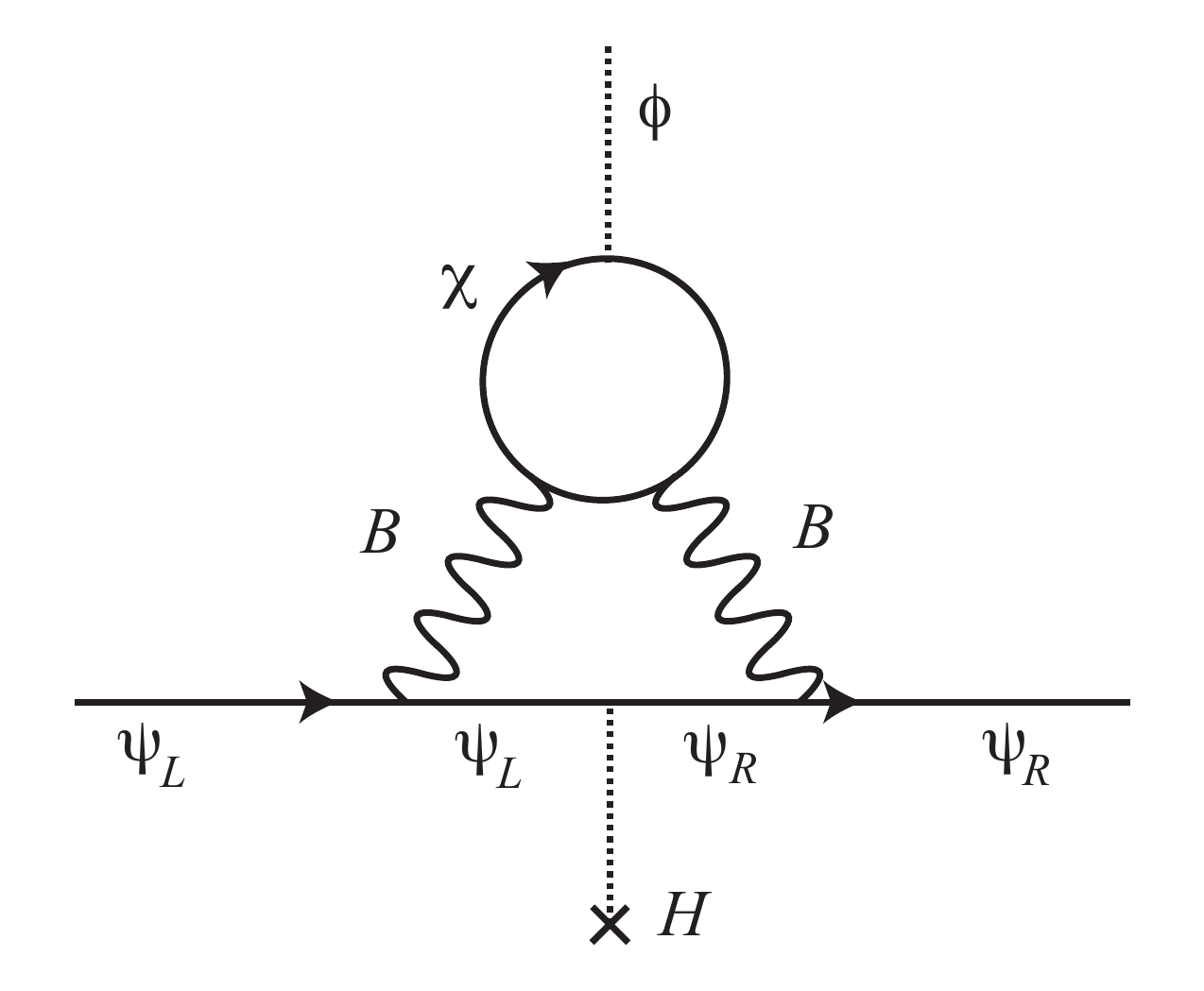}
\includegraphics[height=1.7in, width= 2.5in]{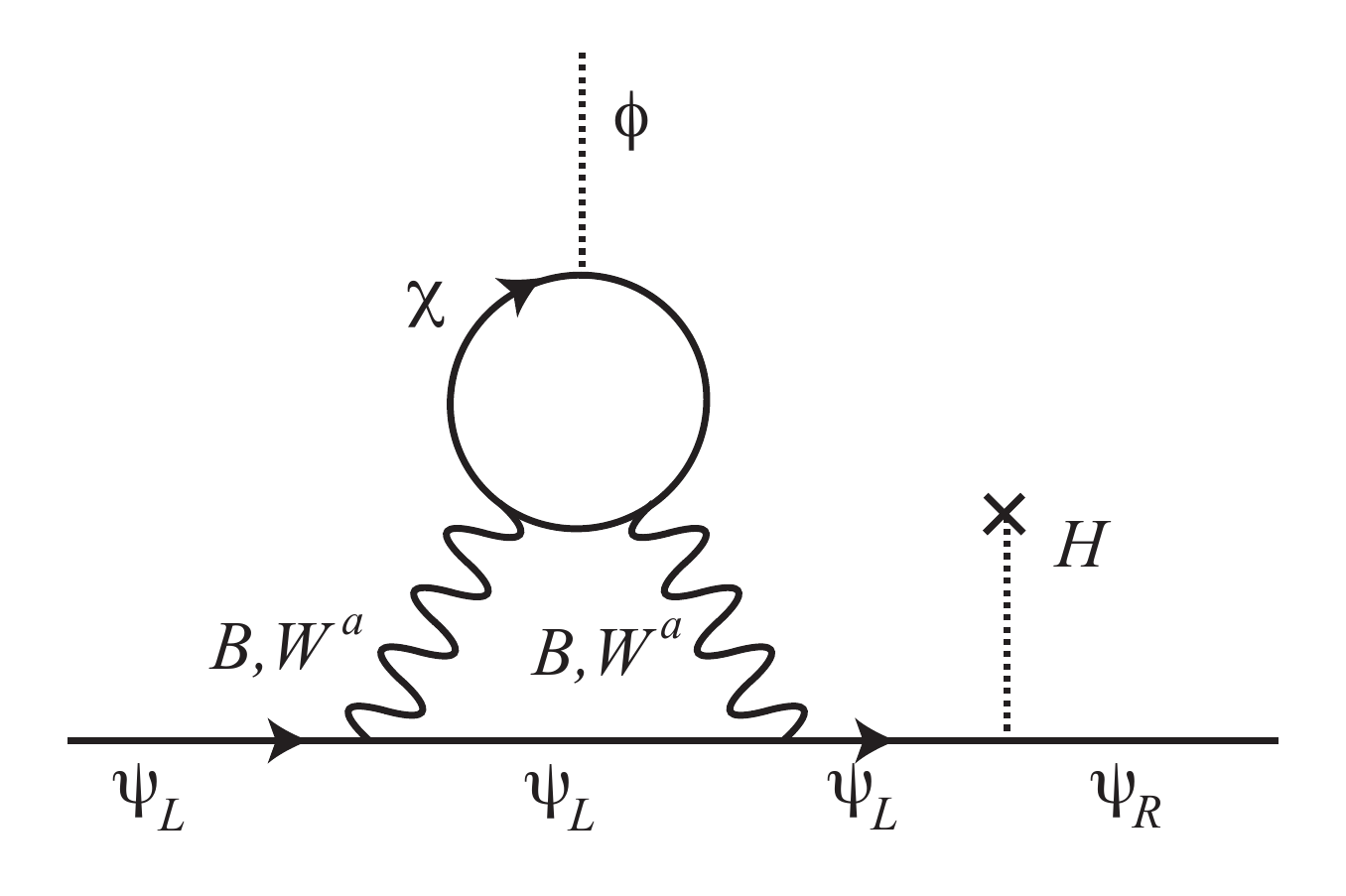}
\caption{Two-loop diagrams in WIMP DM models that generate the operators ${\cal O}^H_{f}$ in Eq.~(\ref{highop}) which generate couplings of the ultralight scalar to the SM fermions after electroweak symmetry breaking. }
\label{gaugeloop}
\end{figure}

\section{WEP Violation Phenomenology}

We first discuss some basic phenomenology of new long range Yukawa forces coupling to ordinary matter or DM. The potential $V$ between two bodies of mass $M_i$ and $M_s$, in the presence of a new long range Yukawa force mediated by a scalar of mass $m_\phi$, can be parameterized as
\bea
  V = -\frac{GM_iM_s}{r}\left (1 + \alpha_{is}e^{-m_\phi r}\right), \qquad   \alpha_{is} = \frac{1}{4\pi G}\frac{Q_iQ_s}{M_iM_s} \xi_i \xi_s,
  \label{alpha}
  \label{totalpotential}
\eea
where the first term in the potential is the usual Newtonian gravitational potential and the the parameter $\alpha_{is}$ is a product of the charge to mass ratios $Q_{i,s}/M_{i,s}$ of the two bodies under the new Yukawa force. The factors of $\xi_{i,s}$ appear for dimensional reasons~\cite{Carroll:2009dw} and are 1 for fermionic test objects and $1/(2M_{i,s})$ for scalar test objects.
The E\"otv\"os parameter $\eta_s$ which measures the relative acceleration between two test objects in the presence of a gravitational source $s$ is given by
\be
  \eta_s = 2\frac{|a_1 - a_2|}{|a_1+a_2|} \simeq \frac{1}{4\pi G}\left|\frac{Q_1 \xi_1}{M_1} - \frac{Q_2 \xi_2}{M_2}\right| \>\left|\frac{Q_s \xi_s}{M_s}\right|,
  \label{eotvos1}
\ee
where we ignored the small mass $m_\phi <10^{-25}$ eV for the ranges $r\ll m_\phi^{-1}$ being probed. The current bounds~\cite{Schlamminger:2007ht} on the E\"otv\"os parameters $\eta_{_E}^{\text{Be,Ti}}$ and $\eta_{_{DM}}^{\text{Be,Ti}}$ which measure the relative acceleration of macroscopic samples of  Beryllium and Titanium  toward the center of the Earth and galactic DM respectively are
\bea
\label{bounds}
 \eta _{_E}^{{Be,Ti}} <  (0.3 \pm 1.8) \times 10^{-13} \qquad  \eta _{_{DM}}^{{Be,Ti}} <  (4 \pm 7) \times 10^{-5}. 
 \eea
 The Apollo (LLR)\cite{Williams:2003wu}  and Microscope\cite{Lammerzahl:2001qr} collaborations are expected to be sensitive to values of $\eta_{_E}^{\text{Be,Ti}}$ as small as $10^{-14}$ and $10^{-15}$ respectively. The MiniSTEP\cite{Lockerbie:1998ar} proposal, if approved, can be sensitive to values of $\eta_{_E}^{\text{Be,Ti}}$ as small as $10^{-18}$.

We parameterize the coupling of the ultralight scalar to DM by 
\bea
\label{eq:chicoup}
\delta {\cal L} =  \begin{cases} 
g_\chi \bar{\chi}  \chi \phi  ,&\text{fermionic DM},\\ 
g_\chi \chi^\dagger \chi \phi,&\text{scalar DM,} \end{cases}
\eea
where $\chi$ denotes the DM field and $g_\chi$ is the coupling of DM to the ultralight scalar. For simplicity we assume that fermionic DM lives in vector-like representations of the electroweak gauge group so that the interaction is gauge invariant. For chiral DM the interaction with the ultralight scalar can arise only from higher dimension operators. From observations in astrophysics~\cite{Kesden:2006vz,Kesden:2006zb} and cosmology~\cite{Bean:2008ac} , the most recent bound on the charge to mass ratio of DM under the new dark force  parameterized by $\beta$ is given as
\bea
\label{eq:beta}
\beta = \frac{M_P}{\sqrt{4 \pi}} \frac{|g_\chi|}{M_\chi} \xi_\chi , \qquad \beta < 0.2,
\eea 
where $M_P$ denotes the Planck Mass. This bound corresponds to a dark force restricted to be less than about 20\% of gravity.
\begin{figure}
\includegraphics[height=2.2in, width = 3in]{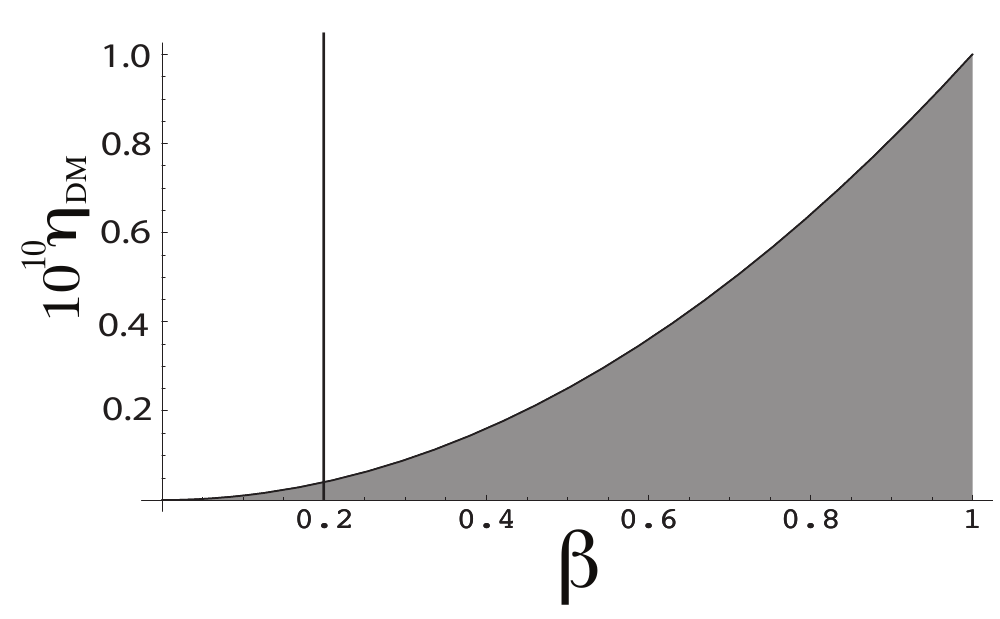} 
\hfill
\includegraphics[height=2.2in, width = 3in]{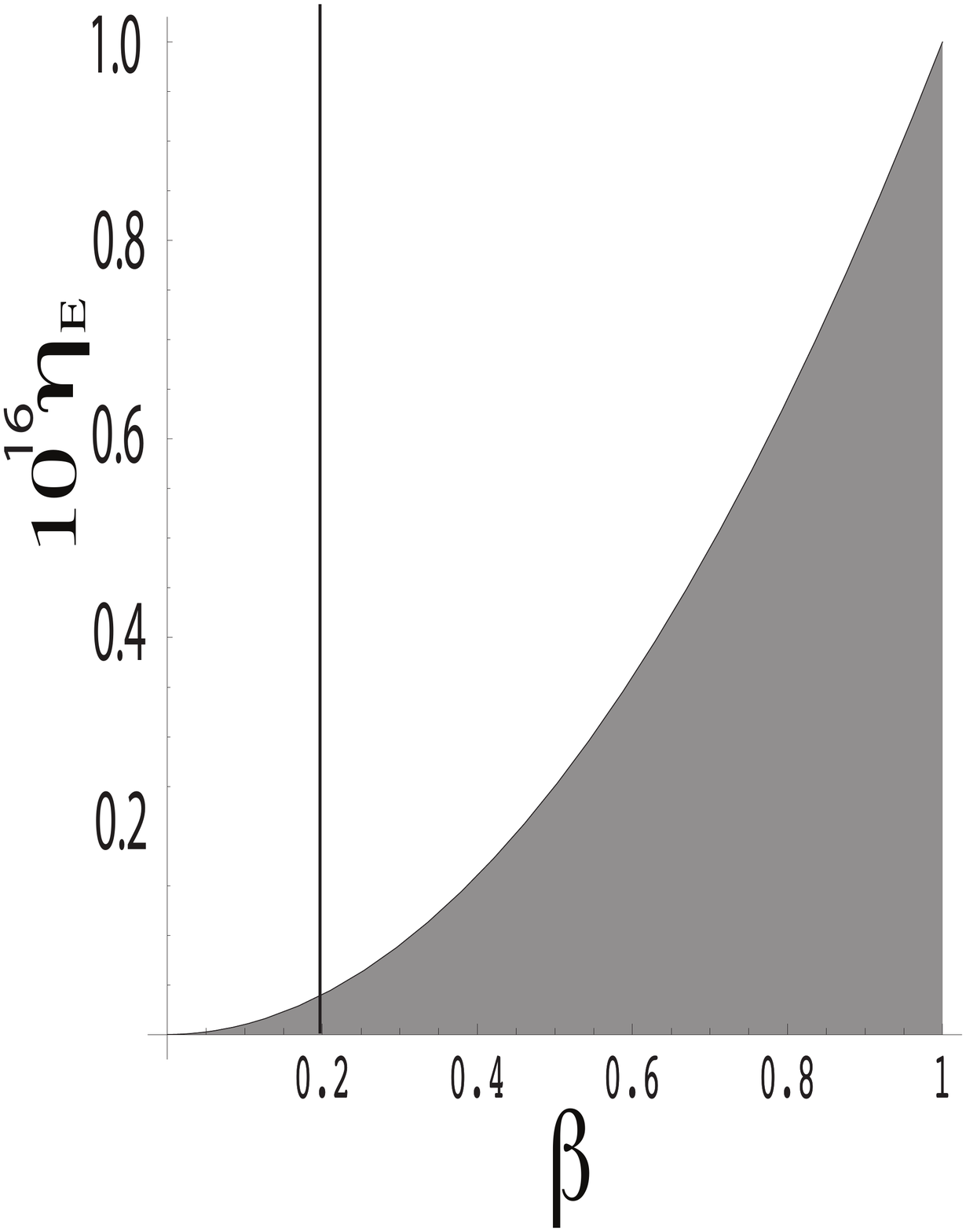}                            
\caption{An estimate of the allowed region in the $(\eta_{_{DM,E}},\beta)$ parameter space for minimal WIMP DM models. The curves in the figures give an estimate of  for $\eta_{_{DM,E}}$ for a given value of $\beta$ from the two loop diagrams in Fig.~\ref{gaugeloop}. The shaded region is unlikely for typical WIMP models. Using the observational constraint $\beta < 0.2$, the allowed region is further restricted to the left of the vertical line. The estimates in the above figures for $\eta_{_{DM,E}}^{\text{Be,Ti}}$, for $\beta < 0.2$, are far below the current experimental bounds $\eta_{_{DM}}^{\text{Be,Ti}} < 10^{-5}, \>\eta_{_{E}}<10^{-13}$. An  improvement of about five orders of magnitude would be required in E\"otv\"os experiments to fully probe the allowed parameter space for $\beta = 0.2$ by measuring $\eta_{_E}^{\text{Be,Ti}}$. This is within reach of the  MiniSTEP proposal\cite{Lockerbie:1998ar} .}
\label{fermionicWIMPetabeta}
\end{figure}

\section{Scalar Dark Forces and Terrestrial Experiments}
\begin{figure}
\includegraphics[height=2in, width=2in]{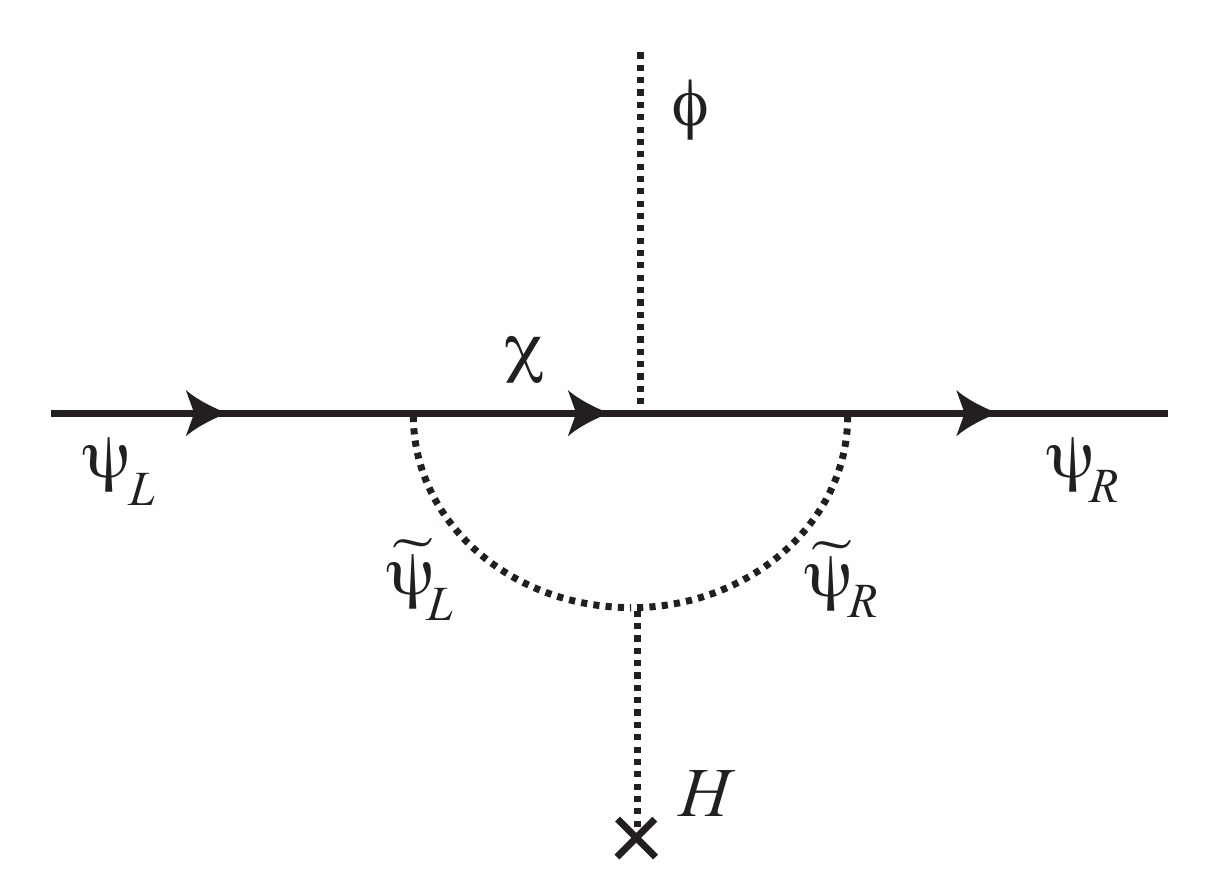}
\caption{DM-induced coupling of $\phi$ to SM fermions at one loop in the presence of additional squark and slepton like degrees of freedom. }
\label{1-loop-squark}
\end{figure}

If the DM has interactions with the SM then the bounds on the parameters $\eta_{_{E,DM}}^{\text{Be,Ti}}$ and $\beta$, obtained from laboratory WEP tests and astrophysical and cosmological observations respectively, will in general  be correlated. As a simple example consider  minimal WIMP DM models ~\cite{Cirelli:2005uq} of the type
\bea
\label{WIMP-lag}
{\cal L} = \begin{cases} 
\bar{\chi}  (i\Dslash + M_0) \chi   , & \text{fermionic DM},\\ 
c  (D_\mu \chi)^\dagger D^\mu \chi  -  c\> M_0^2 \chi^\dagger \chi -V (\chi, H), & \text{scalar DM, } \end{cases}
\eea    
where a single DM particle $\chi$ is added to the SM with non-zero electroweak interactions. Relic density considerations typically require the DM mass to be in the TeV range in such models. We point out that for non-zero hypercharge, such minimal WIMP DM models are typically ruled out~\cite{Cirelli:2005uq} by direct detection experiments. Here we consider such minimal DM models, even with non-zero hypercharge, only as illustrative examples keeping in mind that such DM could be part of a non-minimal extension which avoids the direct detection bounds. If the DM couples directly to a dark force as in Eq.(\ref{eq:chicoup}), then this dark force will couple to the SM quarks and leptons via two diagrams as shown in Fig.~\ref{gaugeloop}. In particular, before electroweak symmetry breaking these loop diagrams generate the dimension five operators of the type
\bea
\label{highop}
{\cal O}_u^H = \phi \bar{Q}_L \epsilon H^\dagger C_u^H u_R, \qquad {\cal O}_d^H = \phi \bar{Q}_L  H C_d^H d_R, \qquad {\cal O}_e^H = \phi \bar{L}_L H C_e^H e_R,
\eea
where the coefficients $C_{u,d,e}^H$ are finite and computable from the loop diagrams. After electroweak symmetry breaking, the Higgs field gets a vacuum expectation value so that these operators generate a coupling between the ultralight scalar and the SM fermions. Note that the coupling of the ultralight scalar to the SM fermions must come from induced  higher dimension operators since no gauge invariant  renormalizable couplings exist between SM fermions and a gauge singlet scalar before electroweak symmetry breaking. From the induced coupling of the ultralight scalar to the SM model fermions one can obtain order of magnitude estimates \cite{Carroll:2009dw} for the E\"otv\"os parameters in terms of the parameter $\beta$.  For minimal WIMP models of the type in Eq.(\ref{WIMP-lag}), the lower bound on the E\"otv\"os parameters $\eta_{_{E,DM}}^{\text{Be,Ti}}$ as a function of $\beta$  is plotted in Fig.~\ref{fermionicWIMPetabeta}.
For $\beta =0.2$, marked by the vertical lines in Fig.~\ref{fermionicWIMPetabeta}, we see that the lower bounds for typical WIMP DM models are $\eta_{_{DM}}^{\text{Be,Ti}} > 4\times 10^{-12}$ and $\eta_{_{E}}^{\text{Be,Ti}}> 4 \times 10^{-18}$. These lower bounds are far below the current experimental upper bounds shown in Eq.(\ref{bounds}). An improvement of about five to seven orders of magnitude in E\"otv\"os experiments would be required in order to probe these lower bounds for WIMP DM models. The MiniSTEP~\cite{Lockerbie:1998ar} experiment, which is currently under study, is expected to reach a sensitivity for $\eta_{_E}^{\text{Be,Ti}}$ of about $10^{-18}$ and might be able to probe the lower bounds.  If an effect is detected in $\eta_{_{DM,E}}^{\text{Be,Ti}}$ far above the lower bounds in Fig.(\ref{fermionicWIMPetabeta}) it would suggest the possibility that the coupling of $\phi$ to the SM fermions might have a different origin. Another possibility that might explain an effect above the expectation in Fig.~\ref{fermionicWIMPetabeta} would be a stronger induced coupling of $\phi$
to ordinary matter arising from non-minimal DM models; for example a one loop coupling of $\phi$ to ordinary matter (see Fig.~\ref{1-loop-squark}) in the presence of additional squark degrees of freedom~\cite{Carroll:2009dw}. The main point is that given a DM model one can estimate the expected size for the E\"otv\"os parameters $\eta_{_{E,DM}}$ for a given value of $\beta$ thus correlating observations in astrophysics and cosmology with laboratory WEP tests.

Another mechanism by which the ultralight scalar  can couple to SM fermions is by mixing with the Higgs. While the ultralight scalar cannot have any renormalizable couplings to the SM fermions before electroweak symmetry breaking, it can have direct couplings to the SM Higgs. After electroweak symmetry breaking, the ultralight scalar will mix with the SM Higgs and in turn  couple to the SM fermions. The size of this mixing angle will be tightly constrained by E\"otv\"os   tests since such an induced coupling to the SM fermions will lead to a violation of the WEP. In what follows we will refer to the ultralight  scalar gauge eigenstate as $S$ and the mass eigenstate after electroweak symmetry breaking as $\phi$.  The renormalizable couplings of the ultralight scalar $S$ to the SM Higgs doublet are encoded in the potential
\bea
\label{Higgs-S}
V (H,S) &=& -\mu^2_h H^\dagger H + \frac{\lambda}{4}  (H^\dagger H)^2  + \frac{\delta _1}{2} H^\dagger H S +
\frac{\delta _2}{2} H^\dagger H S^2 \nn \\
&-& \Big  (\frac{\delta _1 \mu^2_h}{\lambda } \Big ) S + \frac{\kappa_2}{2} S^2 + \frac{\kappa_3}{3}S^3 + \frac{\kappa_4}{4}  S^4.
\eea
We assume for simplicity that $S$ does not acquire a vaccum expectation value.  After electroweak symmetry breaking  the $H^\dag H S$ interaction induces  mixing between the Higgs boson  and the ultralight scalar.  In unitary gauge the neutral component of the Higgs doublet $H$ is given by
\bea
H^0= \frac{v+ h}{\sqrt{2}}, \qquad v=\sqrt{\frac{2\mu_h^2}{\lambda}},
\eea
and the mass terms in the potential become
\bea
\label{Vmass}
V_{\text{mass}} = \frac{1}{2}  (\mu_h^2\> h^2 + \mu_S^2\> S^2 +\mu_{hS}^2 \> h S),
\eea
where
\bea
\label{mu-kappa-delta}
\mu_h^2 = \frac{\lambda v^2}{2}, \qquad \mu_S^2 = \kappa _2 + \frac{\delta _2 v^2}{2}, \qquad \mu^2_{hS} = \delta _1 v.
\eea
The mass eigenstates $h_\pm$ in terms of $S$ and $h$ can be written in terms of a mixing angle $\theta$ as
\bea
\label{mixing}
h_- = S \cos \theta  - h \sin \theta, \qquad h_+ = S\sin \theta  + h \cos \theta, \qquad \tan \theta = \frac{x}{1+ \sqrt{1+x^2}},
\eea
with corresponding masses
\bea
\label{mpm}
m_\pm^2 = \frac{\mu_h^2+\mu_S^2}{2}\pm \frac{\mu_h^2-\mu_S^2}{2} \> \sqrt{1+x^2}, \qquad
x\equiv \frac{\mu_{hS}^2}{\mu_h^2-\mu_S^2}.
\eea

We identify $h_+$ and $m_+$ with the physical Higgs boson field and mass respectively and the physical ultralight scalar field and mass are identified as 
\bea
\label{notation}
\phi \equiv h_-, \qquad m_\phi = m_-. 
\eea
\begin{figure}
\includegraphics[width= 6 in, height =2in]{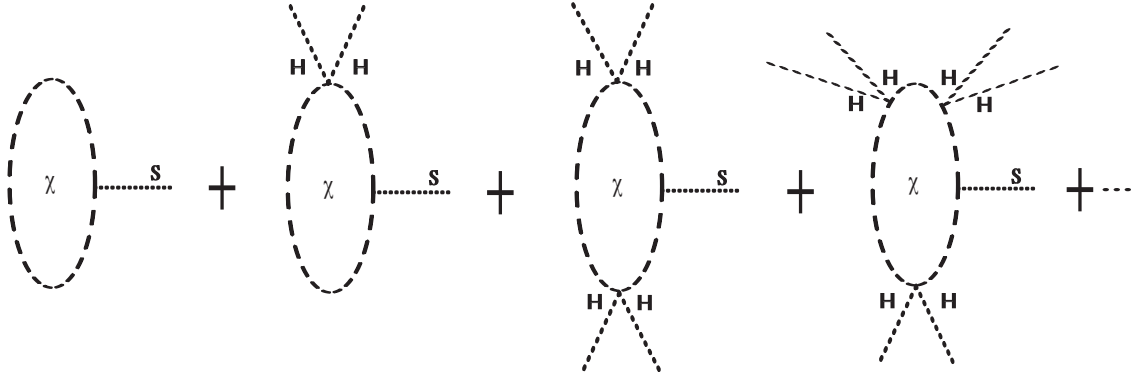}
\caption{One loop diagrams which which contribute to terms in the effective potential $V(H,S)$ that are linear in the $S$ field. The first two diagrams are UV divergent and contribute to the renormalization of the $S$-tadpole and the coupling $\delta_1$ respectively. The remaining diagrams mix into higher dimensional operators and give a finite contribution to Higgs-ultralight-scalar mixing after electroweak symmetry breaking.}
\label{fig:singletmixing}
\end{figure}

Now if the DM couples to the ultralight scalar as in Eq.(\ref{eq:chicoup}),  the mixing angle $\theta$ could  receive additional  contributions from virtual DM effects. As a concrete example consider the DM model where  the DM $\chi$ is either a real singlet or real triplet scalar under the electroweak gauge group with potential
\bea
\label{eq:vhschi}
V(H,S,\chi)&=&V(H,S)+\frac{1}{2}\, M_0^2 \chi^2+\frac{\lambda_\chi}{4}\chi^4+a_2 H^\dag H \chi^2 + g_\chi \chi^2 S + \lambda_{\chi s} \chi^2 S^2, \ \ \ 
\eea
where we have imposed the $Z_2$ symmetry  $\chi \to -\chi$ for the DM field to ensure its stability. The experimental constraints on this DM model in the absence of a dark force  were studied in \cite{Barger:2007im, He:2008qm}. The parameter $a_2$ determines the spin-independent cross-section for DM direct detection
\bea
\label{singlet-cross section}
\sigma_{\chi N} \simeq \frac{a_2^2 \>g_h^2 v^2 m_N^2}{\pi  (M_\chi + m_N)^2 m_h^4}, \qquad M_\chi^2 = M_0^2 + a_2 v^2, \ \ \ 
\eea
where $M_\chi$ is the physical DM mass after electroweak symmetry breaking.
For the singlet DM model, $a_2$ also determines  the relic density as a function of the Higgs mass. We now show that the presence of a dark force can imply an upper bound on the magnitude of  $a_2$ which can translate into an upper bound on the spin-independent direct detection cross-section. For the singlet DM model, the implied upper bound on the magnitude of $a_2$ can be used to rule out dark forces in large regions of parameter space from the DM relic density constraint.

It was shown in \cite{Carroll:2009dw} that  the mixing angle between the ultralight scalar and the Higgs is of the form
\bea
\label{theta-sigma-beta-1}
\sin \theta \approx \tan \theta \approx x \approx    \kappa \frac{a_2 ^2 }{\pi^{3/2}} \frac{ v^3}{M_P m_h^2} \beta + \frac{\delta_1 v}{m_h^2},\ \ \ 
\eea
where $\kappa=1,3$ for the real singlet and real triplet DM respectively.
The second term on the RHS above comes from the renormalizable term $H^\dagger H S$ term after electroweak symmetry breaking.  The first term on the RHS in Eq.(\ref{theta-sigma-beta-1}) is the result of summing the contributions of higher dimension operators of the form $(H^\dagger H)^n S$, for $n\geq 2$, induced via virtual DM as shown in Fig.\ref{fig:singletmixing} at one loop. The first two diagrams in Fig.~\ref{fig:singletmixing} are UV divergent and  just contribute to renormalization of the $S$-tadpole and  the $\delta_1$ coupling respectively. The remaining diagrams are finite and induce operators of the form $(H^\dagger H)^n S$ which give rise to the first term on the RHS of Eq.(\ref{theta-sigma-beta-1})  after electroweak symmetry breaking. For more details see ~\cite{Carroll:2009dw}.
\begin{figure}
\includegraphics[height=2.5in,width=3in]{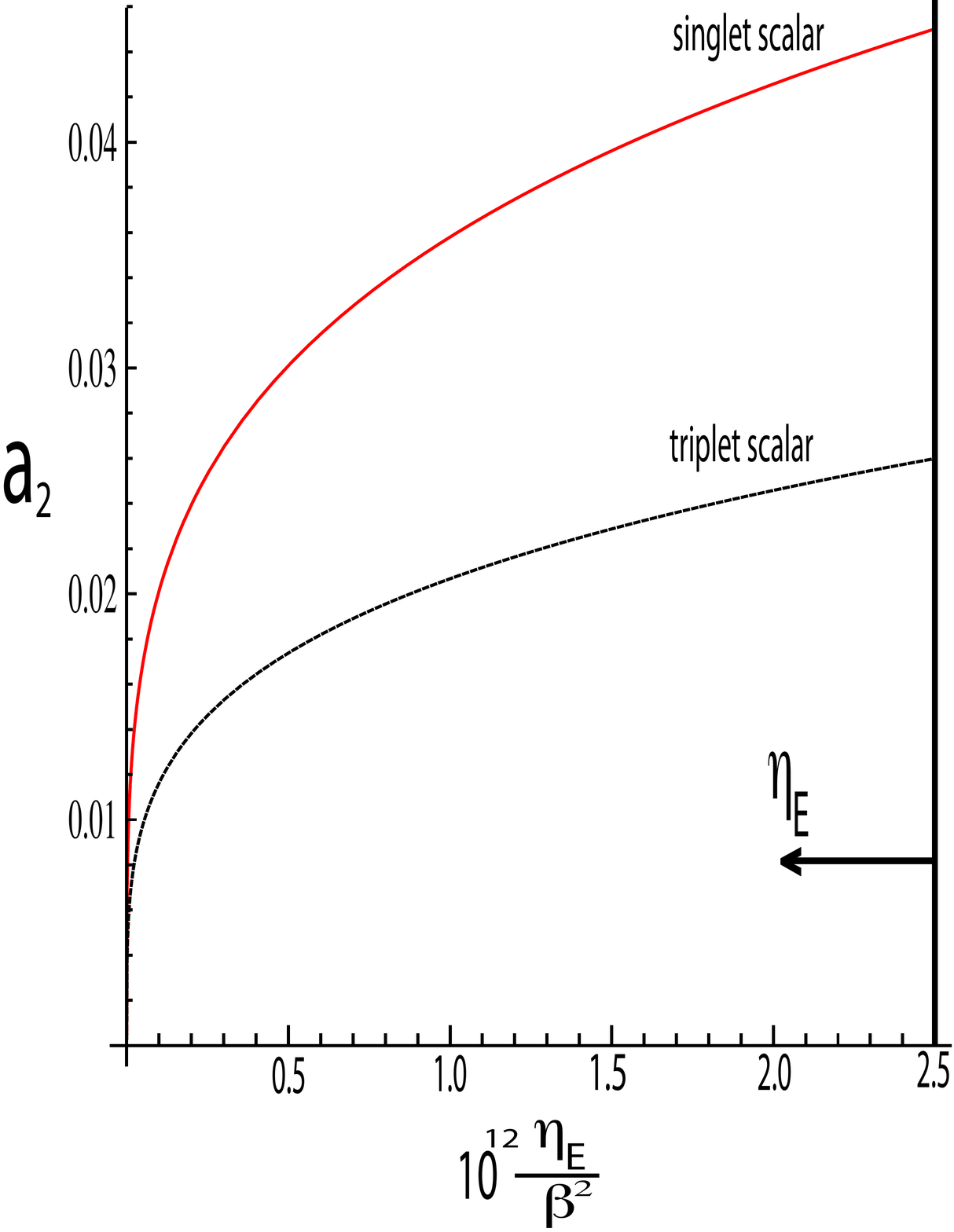}
\hfill \hfill
\includegraphics[height=2.5in,width=3in]{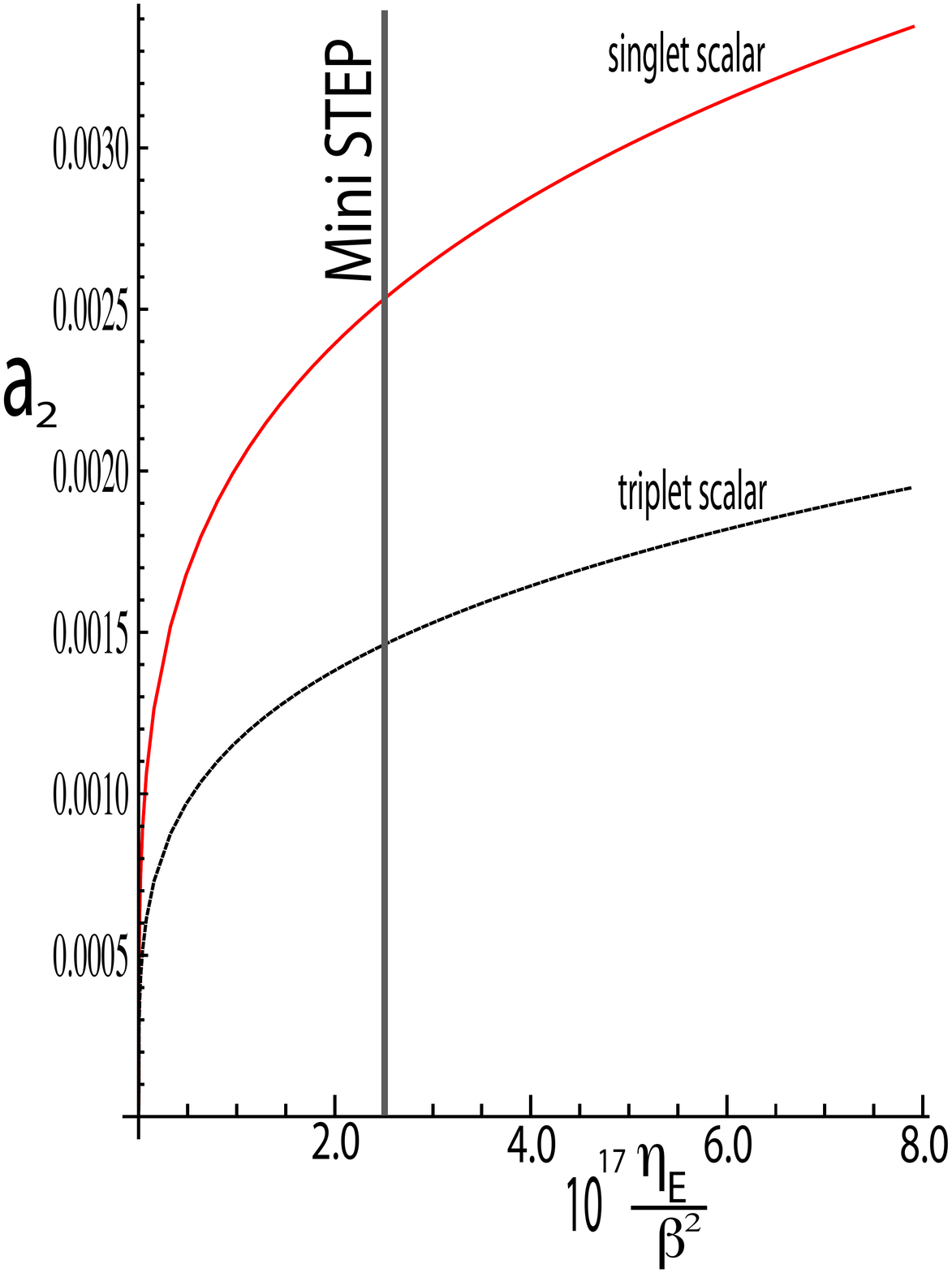}
\caption{Upper bound on $a_2$ in the singlet (upper curve) and real triplet (lower curve) scalar DM models as a function of  $\eta_{_E}^{\text{Be,Ti}}/\beta^2$. The vertical black line in the left panel corresponds to the current bound of $ \eta _{_E}^{{Be,Ti}} < (0.3 \pm 1.8) \times 10^{-13}$. This vertical black line will move to the left with further improvements in E\"otv\"os experiments as indicated by the arrow pointing to the left. The right panel shows the region closer to the expected future bounds from the MiniSTEP experiment and the vertical black line here corresponds to the expected sensitivity of MINISTEP  of $ \eta _{_E}^{{Be,Ti}} = 10^{-18}$.  We have used $m_h=120$ GeV and  $\beta=0.2$. }
\label{a2bound}
\end{figure}

As mentioned earlier, E\"otv\"os experiments will constrain the size of $\sin \theta$ since it leads to WEP violation for ordinary matter and will generate a non-zero $\eta_{_{E,DM}}^{\text{Be,Ti}}$. Furthermore, note that the RHS of Eq.(\ref{theta-sigma-beta-1}) involves the parameter $\beta$ which is constrained from cosmology and astrophysics as in Eq.(\ref{eq:beta}). Except in slices of parameter space where there are intricate cancellations between the two terms on the RHS of Eq.(\ref{theta-sigma-beta-1}), which is unlikely given that $\delta_1$ and $a_2$ are not related in any way, one can use the E\"otv\"os tests to put an upper bound~\cite{Carroll:2009dw}  on the magnitude of $a_2$ as shown in Fig.~\ref{a2bound}.
Figure \ref{a2bound} shows the upper bound on the magnitude of $a_2$ as a function of the ratio $\eta_{_E}^{\text{Be,Ti}}/\beta^2$. In both panels of Fig.~\ref{a2bound}, the upper curve is for the singlet scalar DM model and the lower curve is for the real triplet scalar DM model. The vertical black line on the right end of the  left panel in Fig. \ref{a2bound} corresponds to the values 
 $ \eta _{_E}^{{Be,Ti}} =  (0.3 \pm 1.8) \times 10^{-13}$ and $\beta =0.2$ which are the current bounds. The vertical black line in the right panel in Fig.~\ref{a2bound} corresponds to the value
 $ \eta _{_E}^{{Be,Ti}} =   10^{-18}$, which is the expected sensitivity for the MINISTEP proposal, and $\beta=0.2$. For singlet scalar DM, the upper bound on $a_2$ from Fig.~\ref{a2bound} can be compared with the required value of $a_2$~\cite{He:2008qm} for a given Higgs mass in order to get the right DM relic density and thus further constrain dark forces. For example, for a Higgs mass of 120 GeV and singlet scalar DM mass of 20 GeV, the required value for $a_2$ is about 0.15. In this case a dark force with strength corresponding $\beta=0.2$ is ruled out by the current bound on  $\eta_{_E}^{\text{Be,Ti}}$ shown in Eq.(\ref{bounds}). This can be since in the left panel of Fig.~\ref{a2bound}, which requires $a_2 <0.045$ as seen from the intersection of the upper curve with the vertical line on the right. 
 
\begin{figure}
\includegraphics[width=3in, height=2.5in]{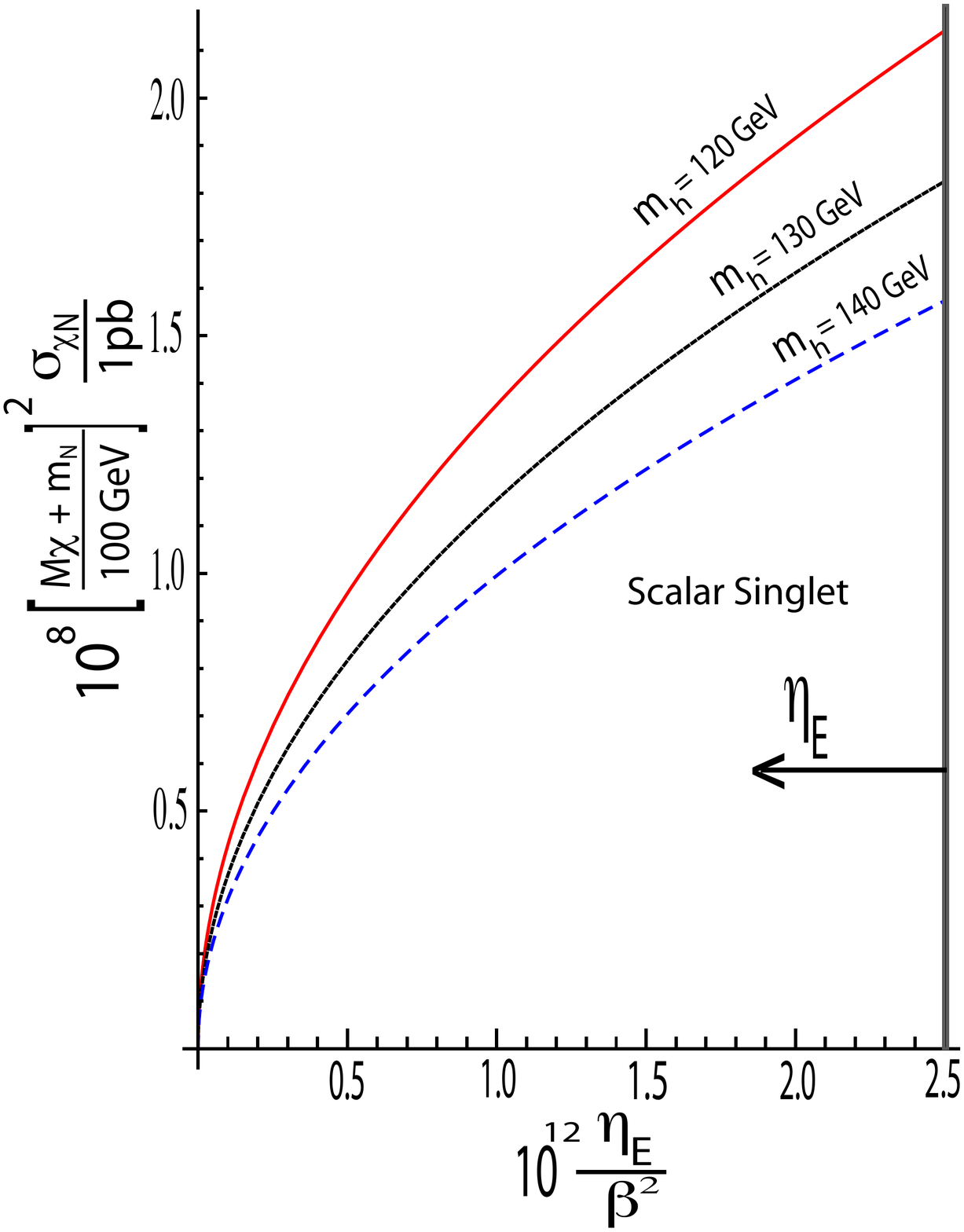}
\hfill
\includegraphics[width=3in, height=2.5in]{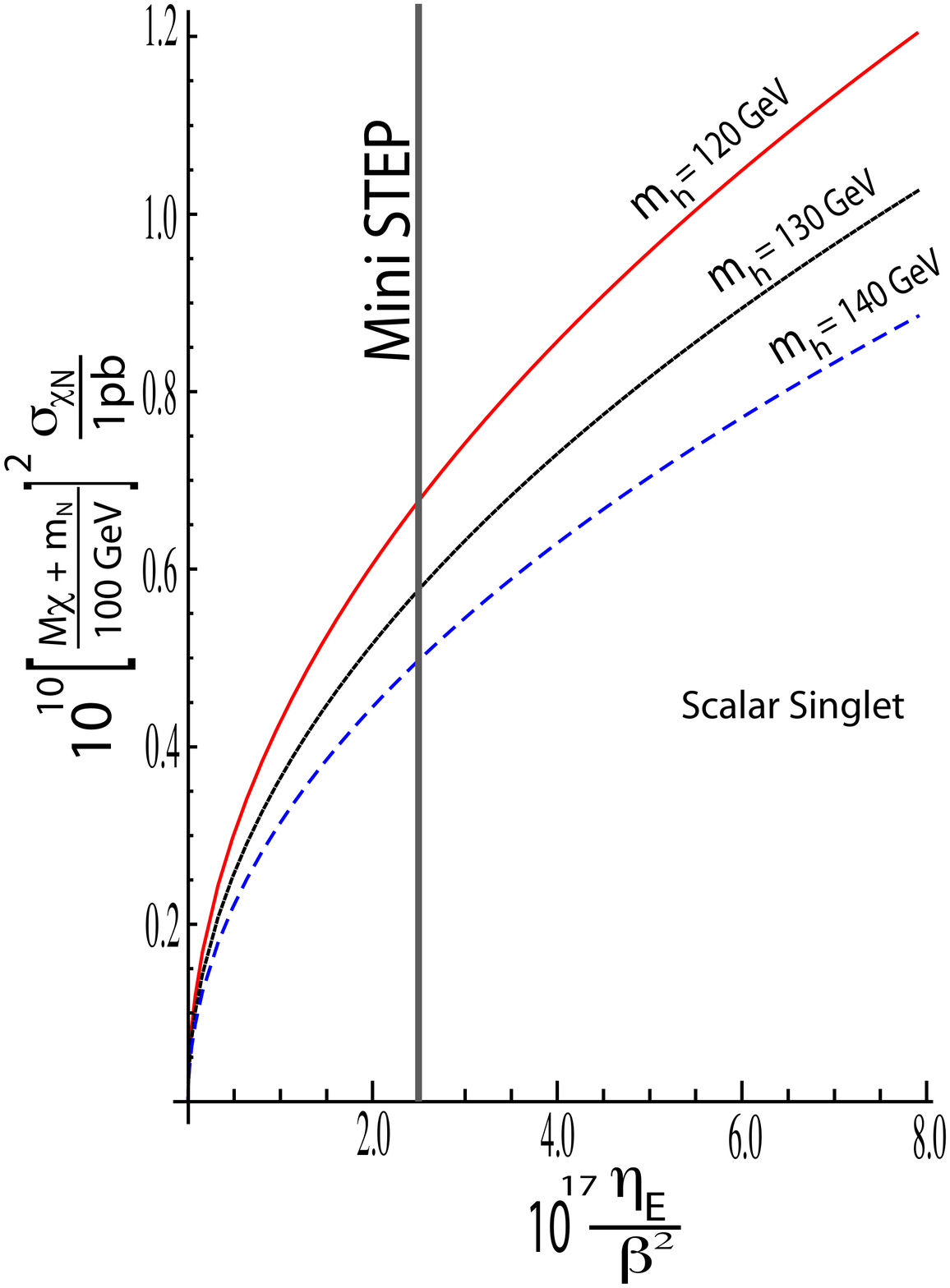}
\caption{Upper bounds on the spin-independent direct detection cross-section for the scalar singlet DM model  as a function of $\eta_{_E}^{\text{Be,Ti}}/\beta^2$. The vertical line in the left panel corresponds to the current bound of $ \eta _{_E}^{{Be,Ti}} < (0.3 \pm 1.8) \times 10^{-13}$. This vertical line will move to the left with further improvements in E\"otv\"os experiments as indicated by the arrow pointing to the left. The right panel shows the region closer to the expected future bounds from the MiniSTEP experiment and the vertical  line here corresponds to the expected sensitivity of MINISTEP  of $ \eta _{_E}^{{Be,Ti}} = 10^{-18}$.  We have used $\beta=0.2$ and the tree curves in each panel correspond to Higgs mass values of $m_h=120,130,140$ GeV as indicated.}
\label{singletWIMPetaE}
\end{figure}

 From Eq.(\ref{singlet-cross section}), the upper bound on the magnitude of $a_2$ also translates into an upper bound on the spin-independent cross-section as shown in Fig.~\ref{singletWIMPetaE} for the singlet DM model. For brevity we have not shown the corresponding plots~\cite{Carroll:2009dw} for the real triplet DM model. 
Figure \ref{singletWIMPetaE} shows how the spin-independent DM direct detection cross-section, E\"otv\"os tests, and observations in astrophysics and cosmology can be brought together to constrain dark forces.  As seen in Fig. \ref{singletWIMPetaE}, smaller values of $\beta$ for a given value of $\eta_{_E}^{\text{Be,Ti}}$, lead to a weaker bound allowing for larger values of the spin-independent direct detection cross-section. As seen from the left panel in Fig.~\ref{singletWIMPetaE}, for a DM mass of around 100 GeV, the upper bounds on the spin-independent cross-section are within reach of current or future sensitivities of direct detection experiments. For a more detailed discussion see ~\cite{Carroll:2009dw}.  Depending on the details of the DM model one can also perform a similar analysis to correlate observations in  WEP tests and astrophysics and cosmology with collider signals. One such example for a real scalar triplet as part of multicomponent DM was studied in ~\cite{Carroll:2009dw}.

As long as the DM is not sterile, the presence of a WEP violating scalar force in the dark sector will be communicated to the ordinary matter sector via virtual DM effects. Depending on the DM model, this effect can be exploited to constrain dark forces weaker than gravity through correlated observations in astrophysics, cosmology, laboratory WEP tests, DM-direct-detection experiments, and collider signals.

\begin{theacknowledgments}

I thank Sean M. Carroll, Michael J. Ramsey-Musolf, and Christopher W. Stubbs for their collaboration that led to the work presented here. This work was supported in part by the U.S. Department of Energy  contract DE-FG02-08ER41531.
\end{theacknowledgments}



\bibliographystyle{aipproc}   

\bibliography{dark}

\IfFileExists{\jobname.bbl}{}
 {\typeout{}
  \typeout{******************************************}
  \typeout{** Please run "bibtex \jobname" to optain}
  \typeout{** the bibliography and then re-run LaTeX}
  \typeout{** twice to fix the references!}
  \typeout{******************************************}
  \typeout{}
 }

\end{document}